\documentclass[twocolumn,showpacs,amsmath,amssymb, showkeys]{revtex4}
\usepackage{graphicx}
\usepackage{dcolumn}
\usepackage{bm}

\begin{document}

\title{Baryonic Tully-Fisher test of Grumiller's modified gravity model}

\author{Samrat Ghosh\thanks{Email address: samrat.ghosh003@gmail.com}} 
\author{Arunava Bhadra\thanks{Email address: aru\_bhadra@yahoo.com}}

\affiliation{High Energy $\&$ Cosmic Ray Research Centre, University of North Bengal, Siliguri, West Bengal, India 734013} 

\author{Amitabha Mukhopadhyay\thanks{Email address:amitabha\_62@rediffmail.com}}
\affiliation{Department of Physics, University of North Bengal, Siliguri, West Bengal, India 734013}

\author {Kabita Sarkar\thanks{Email address: kabita\_id@rediffmail.com}}

\affiliation{Department of Mathematics, Swami Vivekananda Institute of Science $\&$ Technology, Dakshin Gobindapur, Kolkata-700145} 
      
\begin{abstract}
We test the Grumiller's quantum motivated modified gravity model, which at large distances modifies the Newtonian potential and describes the galactic rotation curves of disk galaxies in terms of a Rindler acceleration term without the need of any dark matter, against the baryonic Tully-Fisher feature that relates the total baryonic mass of a galaxy with flat rotation velocity of the galaxy. We estimate the Rindler acceleration parameter from observed baryonic mass versus  rotation velocity data of a sample of sixty galaxies. Grumiller's model is found to describe the observed data reasonably well.
\end{abstract}

\maketitle

Preprint: This a preprint of the Materials accepted for publication in Gravitation and Cosmology, $©$ copyright (2021), the copyright holder indicated in the Journal

\section{Introduction}


Several astrophysical observations and specially the observation of flat rotation curve of galaxies lead to the hypothesis of dark matter. However, despite several efforts so far there is no direct evidence of dark matter particles, nor their existence is predicted by any standard theoretical model of particle physics. Consequently many alternative explanations of flat rotation curve of galaxies exist in the literature including modification of gravitational law at large distances \cite{cap11}, \cite{cli12} or even modification of Newton's laws of dynamics \cite{mil83}. 

Grumiller proposed a quantum motivated theory of gravity that aims to explain the galactic flat rotation in terms of a Rindler acceleration term without the need of any dark matter \cite{gru10}, \cite{gru11}. Assuming spherical symmetry, Grumiller considered the most general form of metric in four dimensions
\begin{eqnarray}
ds^2= g_{\alpha \beta}(x^{\mu})dx^{\alpha} dx^{\beta} + \Phi^2(x^{\mu}) \left(d\theta^2 + sin^2\theta d\varphi^2\right)  \\ \nonumber
,\ \alpha,\beta, \mu =0,1
\end{eqnarray}
where $g_{\alpha \beta}(x^{\mu})$ is a two dimensional metric and the surface radius $\Phi^2(x^{\mu})$ is a 2-dimensional dilaton field. To obtain $g_{\alpha \beta}(x^{\mu})$ and $\Phi^2(x^{\mu})$ Grumiller considered the most general two dimensional renormalizable gravitational theory of the form 

\begin{equation} 
S=-\int\sqrt{-g}[\Phi^{2} R+2{\partial\Phi}^2-6\Lambda\Phi^2+8a\Phi+2]d{^2x}  \, , 
\end{equation}
which contains two fundamental constants, $\Lambda$ and  $a$, the cosmological constant and a Rindler acceleration, respectively. The specialty of the gravitational theory driven by the above action is that it gives a standard Newtonian kind of potential, and the theory has no curvature singularities at large $\Phi(x^{\mu})$. The solution of the two dimensional fields $g_{\alpha \beta}(x^{\mu})$ and $\Phi(x^{\mu})$ are given by

\begin{equation}
g_{\alpha\beta}dx^{\alpha}dx^{\beta} = - B(r) dt^{2} +  \frac{dr^2}{B(r)} ,
\end{equation}

\begin{equation}
 \Phi^2(x^{\mu}) = r^2,
\end{equation}

where
\begin{equation}
B(r) =1- \frac{2M}{r}- \Lambda r^{2} + 2 a r, 
\end{equation}
$M$ is a constant of motion. When $\Lambda= a = 0$, the above solution reduces to the Schwarzschild solution and for $M= \Lambda = 0$ the solution becomes the 2-dimensional Rindler metric. The above solutions are mapped into the four dimensional world through Eq.(1).

The theory has found to explain the rotation curves of spiral galaxies well \cite{lin13}. By fitting the rotation curves of eight galaxies of The HI Nearby Galaxy Survey (THINGS) \cite{wal08} the Rindler acceleration term was found as $a \sim 3 \times 10^{-11}$ $m \; s^{-2}$ {\cite{lin13}. When a larger sample (thirty galaxies) of rotation curves were considered the fitting of the data by the Rindler acceleration was found not very good \cite{mas13}, \cite{cer14} but the goodness of fitting with the Grumiller's theory was still found comparable to that using standard Navarro–Frenk–White (NFW) profile \cite{nav96}, \cite{nav97}. The fitted Rindler acceleration parameter, however, exhibit considerably large spread, at least one order of magnitude with mean around $ 3 \times 10^{-11}$ $m \; s^{-2}$ \cite{cer14}. 

The rotation velocity of galaxies is known to relate with their (galaxies) luminosity \cite{tul77}. The optical Tully-Fisher relation, however, shows break; the relation is not universal for bright and faint galaxies \cite{mcg00}. Instead galactic rotation velocity is found to exhibit universal relation with the total baryonic mass ($M$)  of the galaxy with the form $M \propto v^4_{rot}$ \cite{mcg00}.  

In the present work we have examinedt the Grumiller theory against baryonic Tully-Fisher relation and subsequently we have estimated the Rindler acceleration parameter in the framework of Grumiller's model using observed total baryonic mass versus rotation velocity data for a sample of sixty galaxies. 

\section{Rotation velocity as a function of baryonic matter in Grumiller theory}

For the metric given by Equation (1) with equation (3) the expression of rotation velocity $(v_{rot})$ of galaxies is given by,
 
\begin{equation}
v^2_{rot} =\frac{rB'(r)}{2B(r)}
\end{equation}

where $B'(r)$ signifies the derivative with respect to $r$, $r$ is the co-ordinate distance from galactic centre. For the solution of B(r) given by equation (5) the rotation velocity becomes

\begin{equation}
v^2_{rot} \approx (\frac{m}{r}-\Lambda r^2+ar)^{1/2}
\end{equation}

Because of very small magnitude of $\Lambda$ we henceforth ignore the corresponding term in the expression of rotation velocity.  
The observed rotation velocity in galaxies is in general not strictly constant even at large distances but often has some weak dependence on radial distance. The rotation velocity in Grumiller gravity is also not exactly flat (constant) at large r but slowly increases with r. So an obvious question is what value of rotation velocity will be considered for testing the Tully-Fisher relation. For Grumillers theory we consider (local) minimum value of rotation velocity. The radial distance ($r_e$) at which rotation velocity reaches its extremum value can be obtained by differentiating equation (7) with respect to $r$ and equating it to zero which gives 

\begin{equation}
r_e^2 \simeq \frac{m}{a}
\end{equation}

Inserting it to equation (7), we get,

\begin{equation}
v_{rot}^{\prime 4} = 4am,
\end{equation}

where $v^{\prime}$) denotes the extremum rotation velocity. The above expression shows that Grumiller's theory correctly describe the baryonic Tully-Fisher relation, at least at the theoretical level. 

To match with the observed rotation curve feature a power-law generalization of the Rindler modified Newtonian potential $(-M/r + ar^n)$ is proposed in the literature  \cite{cer14}. Such a power law generalization modifies the Eq.(9) as       

\begin{equation}
v_{rot}^{\prime 4} \propto m^{\frac{n}{n+1}}
\end{equation}

In the above case baryonic mass is not strictly proportional to fourth power of rotation velocity but varies as $m \propto v_{rot}^{\prime 4(n+1)/2n}$.  

\section{Estimation of Rindler acceleration parameter from observed rotation velocity vs Mass data} 

In this section our objective is to estimate the Rindler acceleration parameter from  observed rotation velocity vs baryonic mass data for a sample of disk galaxies. We use the compiled data of Sanders and MacGaugh \cite{san02} as given in Table 1 that include the early works of many good astronomer. 

The major luminous matter components in a typical spiral galaxy are stars and gas. Accordingly the total mass of the galaxy is considered as sum of the stellar mass and gas mass. In the used sample the mass is estimated through photometry, particularly using redder passbands as tracer. The HI thickness method was used for measuring the rotation velocity. The details of the data used and procedure of estimation of mass and rotating velocity are discussed in \cite{mcg00}, \cite{san02}.  

The equation (9) is used to estimate the Rindler acceleration parameter $a$ from the observed data. We fit the observed rotation velocity versus baryonic mass data by the Tully-Fisher relation (equation (9)) using the $\chi^2$ goodness-of-fit test. The fitting gives $a= \left(3.81 \pm 0.01 \right) \times 10^{-11} \; m s^{-2}$ with reduced $\chi^2 =2.0$. The fitted curve is shown in figure 1. 

\begin{figure}[ht]
\begin{center}
\includegraphics[width=0.4\textwidth]{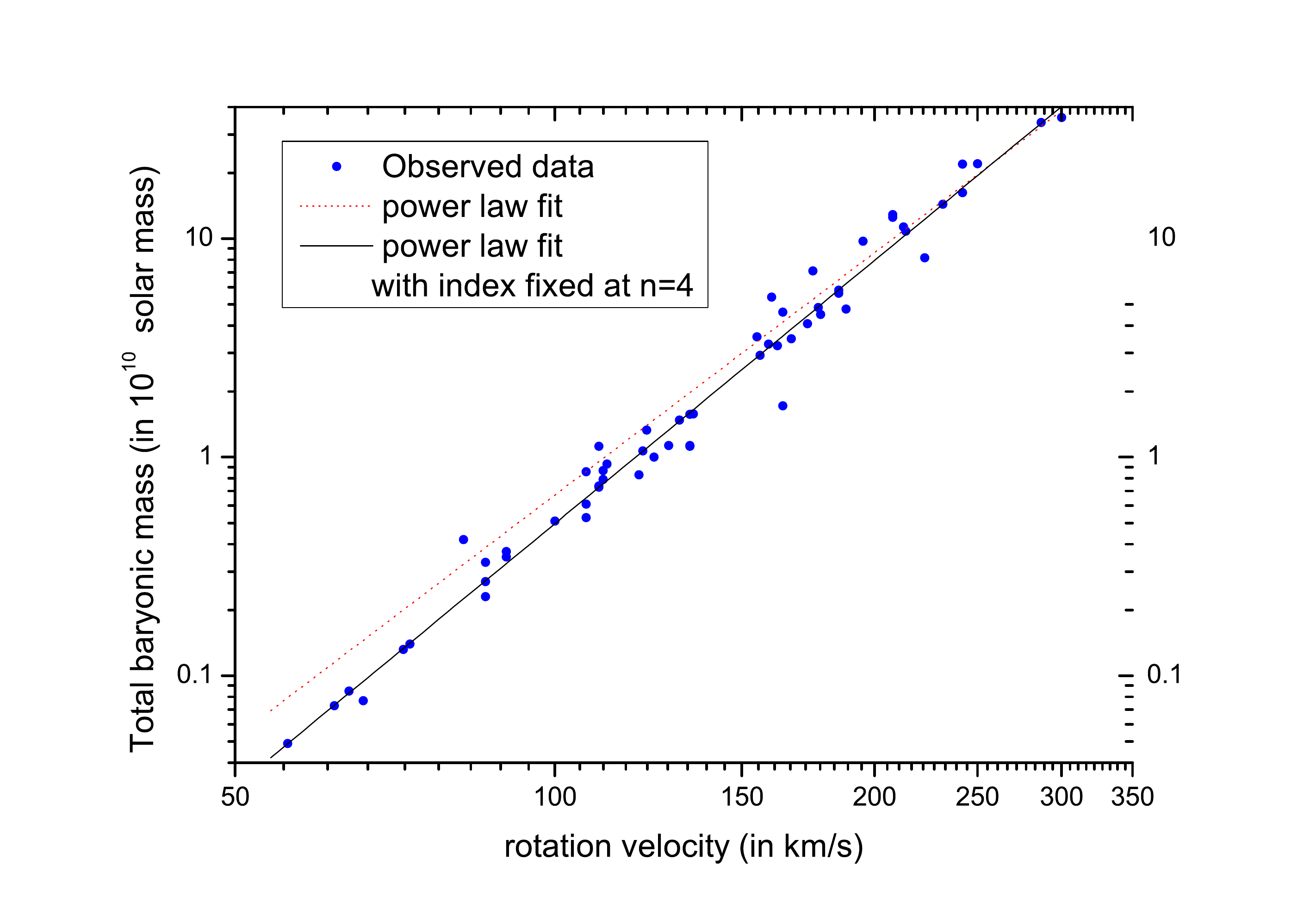}
\end{center}
\caption{Variation of observed total baryonic mass with rotation velocity. The filled (blue) circle represent the observed data, solid (black) line gives the fitting of the data for standard Rindler acceleration (power index fixed at 4) and the dotted (red) line shows the fitting of the data with generalized Rindler acceleration under Grumiller's modified gravity model. }
\end{figure}
   
\begin{table}
\caption{Galaxy data}
\centering
{\scriptsize %
\begin{tabular}{cccccc}
\hline
\hline
\\
Galaxy & $V_{rot}(km \,s^{-1})$ & $M_{stellar}(10^{10} M_{\odot})$ & $M_{gas}(10^{10} M_{\odot})$ & $a (10^{-11}) (m s^{-2})$ \\
\\
\hline
\\
UGC 2885	&300&	30.8	&5&	4.19	\\
NGC 2841	&287&	32.3	&1.7&	3.70	\\
NGC 5533	&250&	19	&3&	3.29	\\
NGC 6674	&242&	18	&3.9&	2.90	\\
NGC 3992	&242&	15.3	&0.92&	3.92	\\
NGC 7331	&232&	13.3	&1.1&	3.73	\\
NGC 3953	&223&	7.9	&0.27&	5.61	\\
NGC 5907	&214&	9.7	&1.1&	3.60	\\
NGC 2998	&213&	8.3	&3&	3.37	\\
NGC 801	&208&	10	&2.9&	2.69	\\
NGC 5371	&208&	11.5	&1&	2.77	\\
NGC 5033	&195&	8.8	&0.93&	2.75	\\
NGC 3893	&188&	4.2	&0.56&	4.86	\\
NGC 4157	&185&	4.83	&0.79&	3.86	\\
NGC 2903	&185&	5.5	&0.31&	3.73	\\
NGC 4217	&178&	4.25	&0.25&	4.13	\\
NGC 4013	&177&	4.55	&0.29&	3.76	\\
NGC 3521	&175&	6.5	&0.63&	2.44	\\
NGC 4088	&173&	3.3	&0.79&	4.06	\\
NGC 3877	&167&	3.35	&0.14&	4.13	\\
NGC 4100	&164&	4.32	&0.3&	2.90	\\
NGC 3949	&164&	1.39	&0.33&	7.79	\\
NGC 3726	&162&	2.62	&0.62&	3.94	\\
NGC 6946	&160&	2.7	&2.7&	2.25	\\
NGC 4051	&159&	3.03	&0.26&	3.60	\\
NGC 3198	&156&	2.3	&0.63&	3.74	\\
NGC 2683	&155&	3.5	&0.05&	3.01	\\
NGC 3917	&135&	1.4	&0.18&	3.89	\\
NGC 4085	&134&	1	&0.13&	5.28	\\
NGC 2403	&134&	1.1	&0.47&	3.80	\\
NGC 3972	&134&	1	&0.12&	5.33	\\
UGC 128	&131&	0.57	&0.91&	3.68	\\
NGC 4010	&128&	0.86	&0.27&	4.40	\\
F568-V1 	&124&	0.66	&0.34&	4.38	\\
NGC 3769	&122&	0.8	&0.53&	3.08	\\
NGC 6503	&121&	0.83	&0.24&	3.71	\\
F568-3 	&120&	0.44	&0.39&	4.63	\\
NGC 4183	&112&	0.59	&0.34&	3.13	\\
F563-V2 	&111&	0.55	&0.32&	3.23	\\
F563-1 	&111&	0.4	&0.39&	3.56	\\
NGC 1003	&110&	0.3	&0.82&	2.42	\\
UGC 6917	&110&	0.54	&0.2&	3.66	\\
UGC 6930	&110&	0.42	&0.31&	3.71	\\
M 33	&107&	0.48	&0.13&	3.98	\\
UGC 6983	&107&	0.57	&0.29&	2.82 	\\
NGC 247	&107&	0.4	&0.13&	4.58	\\
NGC 7793	&100&	0.41	&0.1&	3.63	\\
NGC 300	&90&	0.22	&0.13&	3.47	\\
NGC 5585	&90&	0.12	&0.25&	3.28\\
NGC 55	&86&	0.1	&0.13&	4.40	\\
UGC 6667	&86&	0.25	&0.08&	3.07	\\
UGC 2259	&86&	0.22	&0.05&	3.75	\\
UGC 6446	&82&	0.12	&0.3&	1.99	\\
UGC 6818	&73&	0.04	&0.1&	3.76	\\
NGC 1560	&72&	0.034	&0.098&	3.77	\\
IC 2574	&66&	0.01	&0.067&	4.56	\\
DDO 170	&64&	0.024	&0.061&	3.66	\\
NGC 3109	&62&	0.005	&0.068&	3.75	\\
DDO 154	&56&	0.004	&0.045&	3.72	\\
DDO 168	&54&	0.005	&0.032&	4.26	\\
\hline
\end{tabular}%
}%
\label{Table-1}
\end{table}

The estimated values of $a$ for individual galaxies are given in the last column of the Table 1. It has a small spread, ranges from $1.99 \times 10^{-11} \; m s^{-2}$ for UGC 6446 to $7.79 \times 10^{-11}\; m s^{-2}$ for NGC 3949 with mean value $3.8 \times 10^{-11} \; m s^{-2}$ and standard deviation $0.90$. The frequency distribution of estimated $a$ for the sample of sixty galaxies is shown in figure 2.  

\begin{figure}[ht]
\begin{center}
\includegraphics[width=0.4\textwidth]{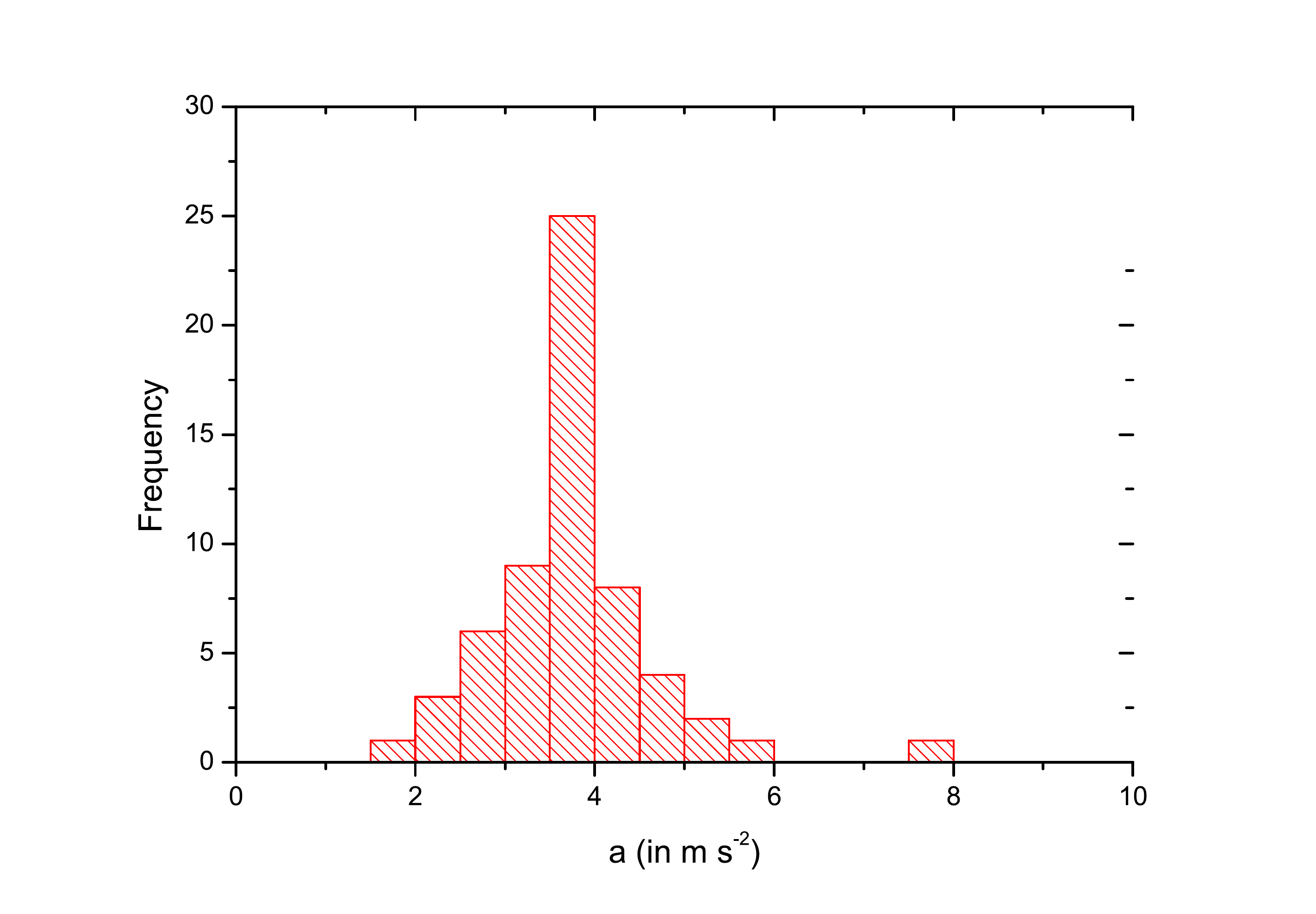}
\end{center}
\caption{Frequency distribution of estimated Rindler acceleration parameter. }
\end{figure}

We also fit the observed rotation velocity versus baryonic mass data for the modified Tully-Fisher relation (equation (10)) led by power-law generalization of the Rindler modified Newtonian potential using the $\chi^2$ goodness-of-fit test which is also depicted in figure 2. In this case the fitted value of the parameters are found $n=1.19$ and $a=9.08 \times 10^{-11} \; m s^{-2}$ with reduced $\chi^2 =1.77$. 

  
\section{Discussion and conclusion}

The Rindler parameter was estimated in \cite{lin13} by fitting rotation curves of eight galaxies of The HI Nearby Galaxy Survey (THINGS) and the fitted mean value of the Rindler acceleration parameter was found $a \sim  3 \times 10^{-11} \; m s^{-2}$. However, when a larger sample of galaxies were considered for analysis the spread in the value of acceleration parameter becomes quite large and thereby the validity of the Grumiller model is questioned \cite{cer14}. In contrast the Rindler acceleration parameter as estimated in the present work using the rotation velocity versus total baryonic mass data of a sample of sixty galaxies exhibits relatively small spread. The mean value is, however, nearly the same to that obtained by fitting rotation curves \cite{cer14}.  As stated already the rotation velocity in Grumiller's theory (equation (7) is not flat but slowly diverges asymptotically which is not in accordance with the observed behaviour in typical rotation curves where rotation velocity is found to decrease slowly at large radial distances \cite{sal07}. This seems the main reason of poor description of rotation velocity curves by the Grumiller's model. While describing the observed rotation velocity versus baryoinc mass data we have considered extrema values of rotation velocity thereby taking out the radial dependence of rotation velocity.                  

It was found in \cite{cer14} that the goodness of fits of rotation curves are better in the generalized Rindler acceleration model than that of the standard Rindler acceleration model. However, the power law index $n$ was found to vary substantially (from $0.2$ to $3.3$) to describe the observer rotation curves \cite{cer14}, which is against the universality of the baryonic Tully-Fisher relation as may be noted from equation (10). The power law generalization is thus not suitable for Tully-Fisher feature unless power law index is kept fixed and universal for all galaxies. Since the form of the Grumiller's solution (Eqs.(3) and (5)) is the same to the vacuum (static spherically symmetric) solution of Weyl gravity \cite{man89}, \cite{kaz91} the present findings are also applicable to Weyl gravity. 
 

The criterion of the stability of orbits in conformal gravity leads to testable upper limit on the size of the galaxies \cite{nan12}. The same conclusion should be applicable to Grumiller's modified gravity because of the similarity of space time solution. Future observations on last stable orbit in galaxies is expected to provide an important test of the Grumiller's model/conformal gravity prediction.  

In conclusion we demonstrate that Grumiller's modified gravity model reproduces the baryonic Tully-Fisher relation at theoretical level if local minimum value of rotation velocity is considered. The fitting of the observed total baryonic mass versus rotation velocity data for a sample of sixty galaxies by Grumiller's model allows to estimate the value of Rindler acceleration parameter. The mean value of so obtained Rindler parameter is found consistent with that estimated from fitting of rotation velocity curves of disk galaxies. 

\begin{acknowledgements} 
AB acknowledge the financial support from DST-SERB, Govt. of India vide approval number CRG/2019/004944.
\end{acknowledgements}

\end{document}